# Physics-Informed Neural Operator for Fast and Scalable Optical Fiber Channel Modelling in Multi-Span Transmission

Yuchen Song[(1)], Danshi Wang*,[(1)], Qirui Fan[(2)], Xiaotian Jiang[(1)], Xiao Luo[(1)], and Min Zhang[(1)]

[(1)] State Key Laboratory of Information Photonics and Optical Communications, Beijing University of Posts and Telecommunications, Beijing 100876, China, *danshi_wang@bupt.edu.cn
[(2)] Department of Electrical Engineering, The Hong Kong Polytechnic University, Hong Kong SAR, China

**Abstract** *We propose efficient modelling of optical fiber channel via NLSE-constrained physics-informed neural operator without reference solutions. This method can be easily scalable for distance, sequence length, launch power, and signal formats, and is implemented for ultra-fast simulations of 16-QAM signal transmission with ASE noise.*

**Introduction**

Nonlinear Schrödinger Equation (NLSE), which does not generally lend itself to analytic solutions, is the basic mathematical model to characterize optical signal propagation in a fiber [1]. The split-step Fourier method (SSFM) is one of the most widely-used methodologies that obtains an approximate numerical solution of the NLSE [2]. However, SSFM is limited in performing large-scale simulation in terms of its high computation complexity and long running time.

To find a closed-form operator that can directly turn the input signals to outputs with fast computation speed has been an overarching goal of fiber modelling. For this purpose, techniques of deep learning were preliminarily explored for optical fiber channel modelling in a data-driven manner, such as bidirectional long short-term memory (BiLSTM) [3] and generative adversarial network (GAN) [4]. Once well trained, the networks act as such an operator and can be faster by up to several orders of magnitude for accurate predictions. However, massive paired input-output data (with reference solutions) are required in the training process, which is time-consuming and cost-prohibitive in data collection.

To utilize the prior knowledge of physics and illuminate the "black-box" of data-driven neural network, physics-informed neural network (PINN) was proposed to solve the partial differential equations (PDEs) directly by embedding physical law-guided PDE into the loss function [5]. PINN has been applied in various scientific and engineering fields [6,7], and recently has also been introduced to optical fiber channel modelling by solving NLSE [8-10]. However, in these works, only simplified optical pulses and low-order optical signal with limited bit lengths were verified in single-span transmissions, which is still far from practical optical links. So far, PINN are used to solve PDE under fixed conditions, and for any given new conditions, it has to retrain another matched network.

To overcome the generalization limitation of PINN, physics-informed neural operator (PINO) [11,12] that blends the benefits of PINN and neural operator [13,14], is lately proposed to learn the solution operator of PDE in the absence of any paired input-output training data.

In this paper, for the first time to our best knowledge, we introduce the framework of PINO to optical fiber channel modelling through learning the solution operator of NLSE. Once the PINO is trained, it can accurately predict outputs at arbitrary transmission distances in a span from new input signals and is easily scalable to different sequence lengths, launch powers and signal formats. We prototype an ultra-fast simulation of 16-QAM multi-span transmission via stacking PINOs with ASE noises, good agreement is observed compared to SSFM.

**Principle of optical fiber channel modelling by PINO**

In mathematical models, transmitted optical signal *s* is complex containing amplitude and phase information and can be expressed as $s=s_I+i\cdot s_Q$ and represented by a vector $[s_I, s_Q]$, where *I* and *Q* denotes the real part (In-phase) and imaginary part (Quadrature), respectively. In the transmission, the signal *s* follows the NLSE:

$$\frac{\partial s}{\partial z}+\frac{\alpha}{2}s+\frac{i\beta_2}{2}\frac{\partial^2 s}{\partial t^2}-i\gamma|s|^2 s = 0 \quad (1)$$

with initial conditions- the input signals, at z=0

$$s(0,\ t) = u(t) \quad (2)$$

where *z* and *t* are input coordinates being the transmission distance and time. Input signals at z=0 can be described by $[u_I, u_Q]$. $\alpha$, $\beta_2$, and $\gamma$ denote attenuation, group velocity dispersion, and nonlinearity coefficient, respectively.

Recently, a novel neural operator architecture coined as DeepONet [13] was proposed and shown to be able to learn the nonlinear operators between Banach spaces [15] with great generalization accuracy by design. Motivated by the PINN, the framework of physics-informed DeepONet [11] was proposed soon afterwards and is adopted in this work. Here, we are interested in learning the solution operator of NLSE $G$ describing mapping from input signals $u(t) \in U$ to the associated solution $s(z, t) \in S$, where $U$ and $S$ are two spaces with operator $G: U \rightarrow S$. The mapping $G_\theta$ is described by $[G_\theta^I, G_\theta^Q]$ with $\theta$ being the parameters to be optimized in the networks. The DeepONet structure consists of two parts of network: the branch net and the trunk net as illustrated in the Fig. 1(b). The trunk net takes the coordinates $(z, t)$, which are generated across the solution domain, as inputs and returns a feature embedding $[k_1, k_2, ..., k_q]^T$. The branch net is further divided into the $I$ and $Q$ part, which both take input signals $u$ as input, where $u = [u(t_1), u(t_2), ..., u(t_m)]$ is evaluated at a collection of fixed locations $\{t_i\}_{i=1}^m$, and output feature embeddings $[b_1, b_2, ..., b_q]^T$. The final signal outputs $[s_I, s_Q]$ are obtained by merging two corresponding feature embeddings by a dot product, which is defined by

$$s^{I(Q)}(z, t) = G_\theta^{I(Q)}(u)(z, t) = \sum_{i=1}^{q} b_i^{I(Q)}\left(u(t_1), u(t_2), ..., u(t_m)\right) k_i(z, t) \quad (3)$$

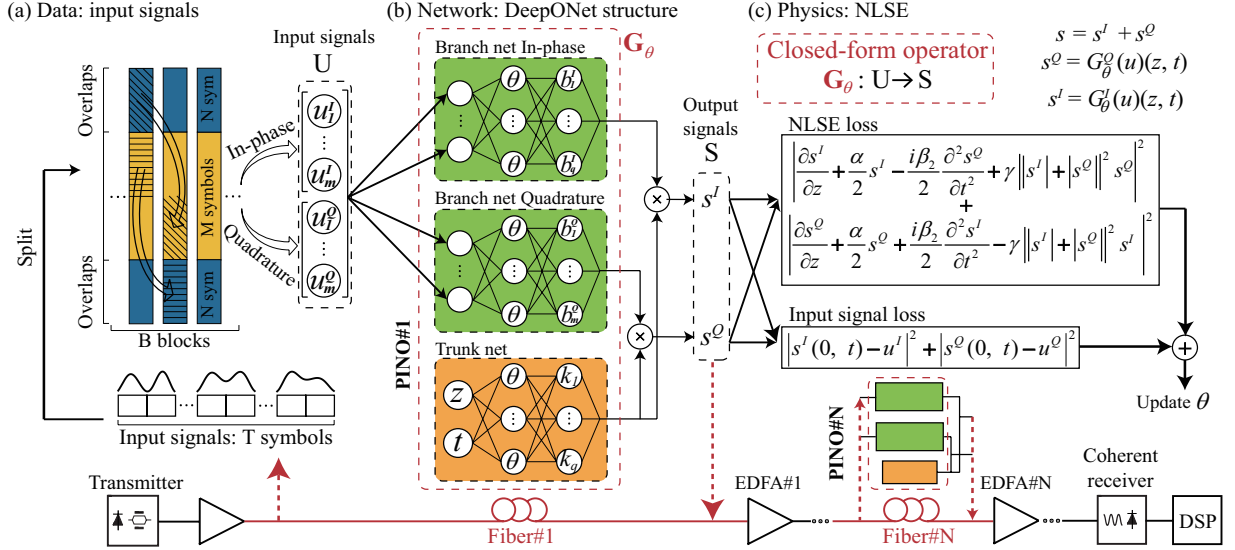

**Fig. 1:** Framework of optical fiber channel modelling by PINO in multi-span transmissions. (a) Processing of input signals, (b) the deep operator network (DeepONet) structure, (c) physics loss functions of NLSE. $\theta$ is the network parameters.

These two outputs are connected to loss functions as shown in Fig. 1(c), where NLSE is embedded and acts as soft constraints to control the outputs under the physical laws. The network parameter $\theta$ is optimized during training. It is worth pointing out that the PINO can predict outputs at different resolutions with the trunk net.

In the training and predicting stage of PINO, we need no paired input-output data, which makes input signals the only known datasets. Due to inter-symbol interference (ISI) caused by chromatic dispersion, a symbol will be affected by its neighbours during the transmission [1], thus overlapping symbols at both ends are required when splitting the long input signal sequence to small ones. As shown in Fig. 1(a), there are N overlapping symbols at both end of each split sequence and M symbols in the middle that can be simulated accurately.

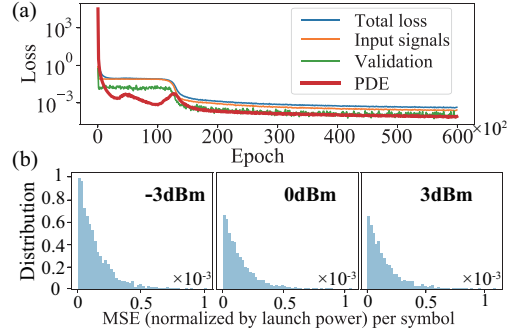

**Fig. 2:** (a) Training losses of PINO, (b) distribution of MSE per-symbol at three launch powers in test data.

Trained in this physics-informed manner, the outputs are not trained to be the labels. Instead, the outputs of the NNs and their derivatives with respect to the inputs rigorously follow the NLSE, which dramatically enhances the generalization of NNs and makes the trained NNs approaching the closed-from solution operator of NLSE.

**16-QAM signal transmission by PINO**

To verify the effectiveness of PINO, 16-ary quadrature amplitude modulation (16-QAM) signals at a line rate of 14 Gbaud to support a net information bit rate of 56 Gbps in a single polarization is studied. Input signals at three launch power levels (-3, 0, 3 dBm) with T = 808 symbols for each are collected after the transmitter with an OSNR of 30dB. Then the long sequence was split into small ones with N = 4 and M = 8. The transmission distance is 80km and the fiber parameters are the same as in [10].

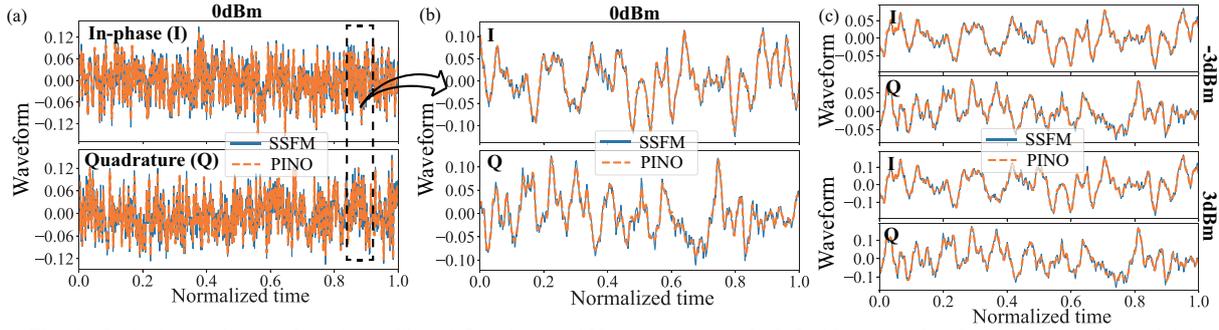

**Fig. 3:** Optical waveforms of In-phase (I) and Quadrature (Q) components of 16-QAM signal after 80-km transmission: (a) 0 dBm launch power; (b) a portion of zoom-in waveforms from (a); (c) -3 dBm and 3 dBm lunch power.

The convergence of losses is recorded in Fig. 2(a). The validation loss calculates the mean square error (MSE) between PINO predictions and SSFM solutions at different transmission distances with new input signals, which proves the good generalization capabilities of the PINO. In the initial stage of training, unlike other losses, the PDE loss is unstable and finally decreases with others, indicating the importance of PDE.

When the PINO is trained, we randomly generate new input signals with launch power at -3, 0, 3 dBm and $T=2^{13}$, M=8, N=4. The distribution of MSE pre-symbol is presented in Fig. 2(b), which indicates more than 95% MSE per-symbol is below $5\times10^{-4}$ for different powers. Corresponding waveforms of in-phase (I) and quadrature (Q) at different launch powers after 80km transmission for SSFM and PINO are compared in Fig. 3, where satisfactory agreement is achieved and proves the solution operator of NLSE is learned with a small collection of input signals (T=808 symbols) and generalizes well for new signal sequence with different launch power levels. It is worthy emphasising that the PINO are trained without any paired input-output data.

To demonstrate our method in a close-to-real optical link, we then transmit signals up to 4 spans with a total of 320km distance through cascading PINOs span by span. The EDFA compensates for the fiber loss with a 5dB noise figure in each span. The PINO#N for the $N^{th}$ span is learned in a recursion fashion. After the PINO#1 for the first span is learned as stated above, the predictions of PINO#1 at 80km were amplified by the EDFA#1 with ASE noises and used as input signals of the span 2. It should be noted that the PINO#2 for the span 2 is learned very quickly based on the parameters of the PINO#1, which applied the framework of transfer learning and leverage the superior generalization ability of the trained PINO. To visualize the intuitive comparison on constellation, both received and compensated constellations by digital backpropagation (DBP) are presented in Fig. 4, where high-fidelity results are obtained by PINO. Compared with SSFM, the PINO has an overwhelming speed advantage, which is about 1,013 times faster at 320km with $2^{17}$ symbols, and the running time do not scale up with longer transmission and more inputs as shown in Fig. 5. Our methods can also obtain good results for signal transmission with low-order formats, such as on-off keying (OOK) and quadrature phase shift keying (QPSK).

**Conclusion**

The closed-form solution operator of NLSE was learned via PINO under the constrain of NLSE without any paired input-output data for optical fiber channel modelling and is demonstrated in 16QAM signal transmission up to four spans. The learned operator can directly tune new input signals to outputs at any transmission distance in a span with a speedup of about 1,000 times compared with SSFM. Thanks to the guidance of NLSE, the trained PINO generalizes well for different input sequences and launch powers and can be easily extended to multiple spans in cascading way with fast fine-tuning. PINO provides a promising scientific computing tool for simulating high-order format signal transmission over multiple spans with additive noise in a fast and low-complexity way.

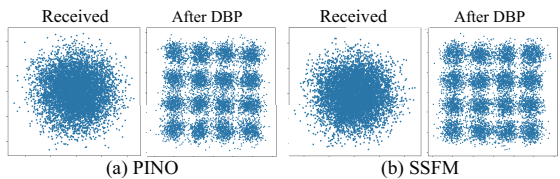

**Fig. 4:** Constellations of 16QAM signal after 320km transmission based on (a) PINO and (b) SSFM

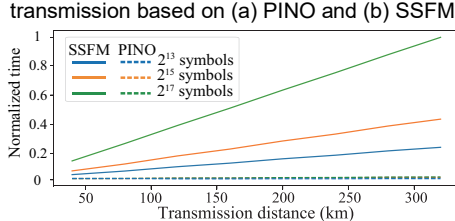

**Fig. 5:** Normalized running time vs. transmission distance of channel model based on SSFM and PINO.

**Acknowledgements**
National Natural Science Foundation of China (No. 62171053, 61975020, 61871415).